# Estimating medical costs from a transition model

## Joseph C. Gardiner[1], Lin Liu[2] and Zhehui Luo[3]

*Michigan State University*

**Abstract:** Nonparametric estimators of the mean total cost have been proposed in a variety of settings. In clinical trials it is generally impractical to follow up patients until all have responded, and therefore censoring of patient outcomes and total cost will occur in practice. We describe a general longitudinal framework in which costs emanate from two streams, during sojourn in health states and in transition from one health state to another. We consider estimation of net present value for expenditures incurred over a finite time horizon from medical cost data that might be incompletely ascertained in some patients. Because patient specific demographic and clinical characteristics would influence total cost, we use a regression model to incorporate covariates. We discuss similarities and differences between our net present value estimator and other widely used estimators of total medical costs. Our model can accommodate heteroscedasticity, skewness and censoring in cost data and provides a flexible approach to analyses of health care cost.

## 1. Introduction

Estimating cost from medical follow-up studies has been the focus of extensive methodological research. Cost data in observational studies exhibit several features such as heteroscedasticity, skewness and censoring that must be addressed in statistical modeling so that ensuing inference would be valid. In clinical trials it is generally impractical to prolong a study until all patients have responded, and therefore inevitably censoring of patient outcomes and total cost will occur in practice. Since costs are incurred over time, the cumulative cost $C(t)$ at time $t$ is a nonnegative monotone function. Cost accumulation ends at an event time $T$, for example at death for lifetime cost, or at a specified finite time horizon $\tau$. Interest lies in estimating the mean cost $\mu = E(C(T^*))$ where $T^* = \min(T, \tau)$. Because $T$ could be precluded from observation by censoring at time $U$, that is, when $T > U$, the corresponding cost would be complete only if $U \geq T^*$. Several nonparametric estimators of $\mu$ have been proposed in a variety of settings with regression models being the mainstay for assessing the influence of patient-specific characteristics (eg, treatments, demographics, comorbidity) on cost (for example, Bang and Tsi-

*Supported by the Agency for Healthcare Research & Quality under grant 1R01 HS14206.
[1]Division of Biostatistics, Department of Epidemiology, B629 West Fee Hall, Michigan State University, East Lansing, MI 48824, USA, e-mail: jgardiner@epi.msu.edu
[2]Now at Eli Lilly and Company, US Medical Division, Indianapolis, Indiana 46285, USA, e-mail: liu_lin_ll@lilly.com
[3]Now at RTI International, Behavioral Health Economics Program, 3040 Cornwallis Rd., Research Triangle Park, NC 27709, USA, e-mail: zluo@rti.org

atis [2, 3], Baser et al. [4, 5], Lin [13, 14], Lin et al. [15], O'Hagan and Stevens [17], Strawderman [18] and Gardiner et al. [8]).

This article adopts a broader view of the cumulative cost $\{C(t) : 0 \leq t \leq \tau\}$ within the framework of a longitudinal model. Section 2 describes all the substantive aspects of our models starting with an underlying finite state stochastic process for the evolution of patient events as they occur over time. The states are different health conditions that the patient presents over the period $[0, \tau]$. Costs emanate from two streams, during sojourn in health states and in transition from one health state to another. We consider estimation of net present value (NPV) for expenditures incurred over $[0, \tau]$. Regression models for the event history process and for observed costs are used to incorporate covariates. Section 3 outlines the method of estimation of NPV from a patient sample of time-censored event history data. We then discuss similarities and differences between our net present value estimator and other widely used estimators of total medical costs. Section 4 is a brief summary and conclusion.

## 2. Stochastic model

### *2.1. Transition and sojourn cost*

A stochastic process $X = \{X(t) : t \in \mathcal{T}\}$ on the interval $\mathcal{T} = [0, \tau]$ where $\tau < \infty$, describes the health states of a patient from the relevant population under study. The time $\tau$ is the maximum limit of observation for all cost and patient outcomes. The state space of $X$ is finite and labeled E= $\{0, \ldots, m\}$ and consists of several transient states, such as "well", "recovery", "relapse", and one or more absorbing states such as "dead" or "disabled". A transient state is one which if visited will be exited after a finite sojourn, whereas a transition out of an absorbing state is impossible. Costs are incurred while sojourning in a transient health state and in transition between states. If the patient is in state $h$ at time $t$, that is, $X(t) = h$, the expenditure rate is $B(t, h)$. If a transition occurs from state $h$ to state $j$ at time $t$, that is, $X(t-) = h$ and $X(t) = j$, a cost $C(t, h, j)$ is incurred.

The notation $[A]$ denotes the indicator function of the event $A$ taking value 1 if $A$ is true and 0 if $A$ is false. For example, to indicate the state of occupation just prior to time $t$ we write $Y_h(t) = [X(t-) = h]$. The number of direct transitions $h \to j$, $h \neq j$ in the time interval $[0, t]$ is $N_{hj}(t) = \#\{s \leq t : X(s-) = h, X(s) = j\}$. If $r$ is the discount rate, the present value of expenditures associated with all $h \to j$ transitions in $\mathcal{T}$ is

$$C_{hj}^{(1)} = \int_0^\tau e^{-rt} C(t, h, j) dN_{hj}(t), \tag{1}$$

and the present value of expenditures for all sojourns in state $h$ in $\mathcal{T}$ is

$$C_h^{(2)} = \int_0^\tau e^{-rt} B(t, h) Y_h(t) dt. \tag{2}$$

We will interpret all integrals as on the semi-open interval $(0, \tau]$. In practice we want to estimate the expected values (averages) of these two quantities. To do so we impose a Markov model on $X$ to govern the transitions between states.

### 2.2. Markov model

We call $X$ a *non-homogeneous Markov process* if

$$P[X(t) = j \mid X(s) = h, X(u) : u < s] = P[X(t) = j \mid X(s) = h]$$

for all $h, j \in$ E and all $s \leq t$. The *transition probabilities* $P_{hj}(s,t)$, $s \leq t$, of $X$ are given by $P_{hj}(s,t) = P[X(t) = j \mid X(s) = h]$ and the *transition intensities* $\alpha_{hj}(t)$ by $\alpha_{hj}(t) = \lim_{\Delta t \downarrow 0} P[X(t + \Delta t) = j \mid X(t) = h]/\Delta t$, $j \neq h$ with $\alpha_{hh} = -\sum_{j \neq h} \alpha_{hj}$. Throughout we assume that the $\alpha_{hj}$ are integrable on $\mathcal{T} = [0, \tau]$. The $m \times m$ matrices $\mathbf{P} = \{P_{hj}(s,t)\}$ and $\boldsymbol{\alpha} = \{\alpha_{hj}\}$ are related by the product-integral formula $\mathbf{P}(s,t) = \prod_{s < u \leq t} (\mathbf{I} + \boldsymbol{\alpha}(u)du)$. The matrix $\mathbf{A} = \{A_{hj}\}$ of the integrated intensities is defined by $A_{hj}(t) = \int_0^t \alpha_{hj}(u)du$.

### 2.3. Modeling covariates

We let the transition intensities depend on a covariate history vector process $\mathbf{z}(t)$ through a Cox regression model $\alpha_{hj}(t|\mathbf{z}(t)) = \alpha_{hj0}(t) \exp(\beta'_{hj} \mathbf{z}(t))$, where $\alpha_{hj0}(t)$ is an unknown baseline intensity and the regression coefficients $\beta_{hj}$ are specific to the transition $h \to j$. It is always possible to recast this in terms of a single composite regression vector $\beta$ with type-specific covariate vector $\mathbf{z}_{hj}(t)$ computed from $\mathbf{z}(t)$. Then the model for the intensities is

$$\alpha_{hj}(t|\mathbf{z}(t)) = \alpha_{hj0}(t) \exp(\beta' \mathbf{z}_{hj}(t)). \tag{3}$$

To make explicit the dependence of $\mathbf{P}$, $\boldsymbol{\alpha}$ and $\mathbf{A}$ on a pre-specified fixed covariate profile $\mathbf{z}$ we will use the notation $\mathbf{P}(s,t|\mathbf{z})$, $\boldsymbol{\alpha}(t|\mathbf{z})$, and $\mathbf{A}(t|\mathbf{z})$, respectively.

### 2.4. Net present value

Consider the conditional expectation of (1), given fixed $\mathbf{z}$ and the initial state $X(0) = i$, $i \in$ E. The expected net present value is

$$E(C_{hj}^{(1)} | X(0) = i, \mathbf{z}) = \int_0^\tau e^{-rt} c_{hj}(t|\mathbf{z}) P_{ih}(0, t-|\mathbf{z}) dA_{hj}(t|\mathbf{z}), \tag{4}$$

where $c_{hj}(t|\mathbf{z}) = E\{C(t, h, j) | X(t-) = h, \mathbf{z}\}$ and integration is on the set $(0, \tau]$. Justification for (4) could be made as follows. Starting in state $i$ at time zero a patient will be in state $h$ at time $t$ with probability $P_{ih}(0, t|\mathbf{z})$. Conditional on being in state $h$ just prior to $t$, a transition to state $j$ occurs at $t$ with intensity $\alpha_{hj}(t|\mathbf{z})$ and this transition incurs a cost whose average is $c_{hj}(t|\mathbf{z})$. We call (4) the net present value (NPV) for all $h \to j$ transition costs in $\mathcal{T}$. An entirely analogous argument applies to the conditional expectation of (2), given $\mathbf{z}$ and the initial state $X(0) = i$, which results in the NPV for all sojourn costs in state $h$ in $\mathcal{T}$. We get

$$E(C_h^{(2)} | X(0) = i, \mathbf{z}) = \int_0^\tau e^{-rt} b_h(t|\mathbf{z}) P_{ih}(0, t-|\mathbf{z}) dt, \tag{5}$$

where $b_h(t|\mathbf{z}) = E\{B(t, h) | X(t-) = h, \mathbf{z}\}$. The interpretation of (5) is as follows. Starting in state $i$ at time zero a patient will be in state $h$ just prior to time $t$ with probability $P_{ih}(0, t-|\mathbf{z})$. While sojourning in state $h$ in the interval $(t, t + dt]$ an average cost $b_h(t|\mathbf{z})dt$ is incurred. Then the right hand side of (5) is a weighted sum of these costs in $[0, \tau]$.

By averaging with respect to the initial distribution, $\pi_i(0|\mathbf{z}) = P[X(0) = i|\mathbf{z}]$, $i \in E$ we combine (4) and (5) to obtain the unconditional NPV,

$$\text{NPV}(\mathbf{z}) = \sum_{i \in E} \pi_i(0|\mathbf{z})\text{NPV}(\mathbf{z}, i),$$

where

(6) $$\text{NPV}(\mathbf{z}, i) = \sum_{h \neq j} \int_0^\tau e^{-rt} c_{hj}(t|\mathbf{z}) P_{ih}(0, t-|\mathbf{z}) dA_{hj}(t|\mathbf{z})$$
$$+ \sum_h \int_0^\tau e^{-rt} b_h(t|\mathbf{z}) P_{ih}(0, t-|\mathbf{z}) dt.$$

Since there is no cost accumulation in absorbing states (e.g., no costs incurred after death) the second summation is over all transient states in E. The first summation includes both transitions between transient states and transitions from a transient state to an absorbing state. We can simplify (6) further by defining $c_h^*(t|\mathbf{z}) = \sum_{j \neq h} c_{hj}(t|\mathbf{z}) \alpha_{hj}(t|\mathbf{z})$ and rewriting the first term as

$$\sum_h \int_0^\tau e^{-rt} c_h^*(t|\mathbf{z}) P_{ih}(0, t-|\mathbf{z}) dt.$$

This is similar to the second term in (6).

Equations (4)-(6) place a structure to the accumulating costs by considering costs incurred at transitions separately from costs incurred during sojourns. In general, we might consider two non-negative, non-decreasing, right-continuous processes $\{V_k(t) : t \in \mathcal{T}, k = 1, 2\}$ to represent the cost accumulation which is assumed to end at time $\tau$, or prior to $\tau$ if an absorbing state has been entered. For costs incurred at transition times, $V_1(t) = \sum_{h \neq j} \sum_{u \leq t} e^{-ru} C(u, h, j) \Delta N_{hj}(u)$, and for costs incurred during sojourn in states we have $V_2(t) = \sum_h \int_0^t e^{-ru} B(u, h) Y_h(u) du$.

## 2.5. *Censoring*

Observation of $X$ will cease at time $\tau$ unless an absorbing state was entered prior to $\tau$. Also censoring might occur at some random time $U$, which limits observation up to $\tau \wedge U$. We assume $U$ is independent of $X$ and replace $N_{hj}(t)$ by the censored process $N_{hj}(t \wedge U)$ and the state indicator $Y_h(t)$ by $Y_h(t) = [X(t-) = h, U \geq t]$. Therefore, for the process $X$ the information $\mathcal{F}_t$ revealed up to time $t$ is generated from $X(0), \mathbf{z}(0)$, and $\{\mathbf{z}(u), Y_h(u), N_{hj}(u) : u \leq t \wedge U, h \neq j, h, j \in E\}$. For costs incurred at transition times the information known up to time $t$ is

$$\{C_{hj}(u) \triangle N_{hj}(u) : u \leq t \wedge U, h \neq j, h, j \in E\},$$

whereas for sojourns we would know at best the cumulative costs

$$\{\int_0^{t \wedge U} B_h(u) Y_h(u) du, h \in E\}.$$

In both cases censoring limits what we can observe. If $U$ precedes both $\tau$ and the time of absorption then total costs are not observed. Furthermore, the observational scheme might restrict observation of sojourn costs to only completed sojourns or at a finite number of time points during the sojourn. We assume that censoring completely random in the sense that $U$ is independent of $(X(t), V(t) : t \geq 0)$.

## 2.6. Survival time

Label the states in E so that $0, \ldots, m-1$ are transient and $m$ is absorbing (e.g., the state 'dead'). Survival time $\tau_m = \inf\{t > 0 : X(t) = m\}$ is the time to absorption. The survival distribution, conditional on $X(0) = i$, is

$$S_{mi}(t|\mathbf{z}) = P[\tau_m > t | X(0) = i, \mathbf{z}] = 1 - P_{im}(0, t|\mathbf{z}),$$

and the unconditional survival distribution is

$$S_m(t|\mathbf{z}) = 1 - \sum_{i \neq m} \pi_i(0|\mathbf{z}) P_{im}(0, t|\mathbf{z}).$$

In the special case of one transient state 0 ("alive") and one terminal state 1 ("dead") we get the usual survival time $T(= \tau_1)$ and its survival distribution $S(t|\mathbf{z}) = P[T > t|\mathbf{z}] = P_{00}(0, t|\mathbf{z})$.

## 3. Estimation

Suppose we observe the aforementioned processes for each of $n$ subjects in a longitudinal study. For the $i$-th patient the basic covariate vector is $\mathbf{z}_i(t)$, the initial state $X_i(0)$, the state indicator $Y_{hi}(t) = [X_i(t-) = h, U_i \geq t]$ and the number of direct $h \to j$ transitions

$$N_{hji}(t) = \#\{u \leq t \wedge U_i : X_i(u-) = h, X_i(u) = j\}, h \neq j.$$

Conditionally on $\{\mathbf{z}_i(0), X_i(0) : 1 \leq i \leq n\}$ assume processes $\{X_i(t) : t \in \mathcal{T}\}$ are independent and that model (3) holds for each individual with the same baseline intensities. From now on denote by $N_{hj}(t)$ and $Y_h(t)$, respectively, the aggregated processes $\sum_{i=1}^{n} N_{hji}(t)$ and $\sum_{i=1}^{n} Y_{hi}(t)$. In this context estimation of the transition probabilities $P_{hj}(0, t|\mathbf{z})$ and integrated intensities $A_{hj}(t|\mathbf{z})$ at a fixed covariate profile $\mathbf{z}$ is well known (Andersen et al. [1]). Combining this with appropriate estimation of costs would lead to estimators of NPV. However, before we describe an approach to estimation we first consider several examples.

### 3.1. Single transition without covariates

The only permissible transition $0 \to 1$ is associated with a single cost $C(T, 0, 1)$ (denoted here by $y$) where $T$ denotes the survival time. From (6) we have

$$\text{NPV} = \int_0^{\tau} e^{-rt} c_{01}(t) P_{00}(0, t-) dA_{01}(t).$$

To estimate NPV we use the estimators

$$\hat{P}_{00}(0, t-) = \hat{S}(t-)$$

and

$$d\hat{A}_{01}(t) = \{Y_0(t)\}^{-1} dN_{01}(t),$$

where $\hat{S}$ is the Kaplan-Meier estimator of the survival distribution of the survival time $T(= \tau_1)$, $Y_0(t) = \sum_{i=1}^{n}[T_i \wedge U_i \geq t]$ and $N_{01}(t) = \sum_{i=1}^{n}[T_i \leq t \wedge U_i]$. If there are no ties in the survival times $T_i$, the natural estimator of $c_{01}$ is $\hat{c}_{01}(T_i) = y_i$. Therefore our estimator of NPV is

$$\text{(7)} \quad \hat{\text{NPV}} = \int_0^\tau e^{-rt} \hat{c}_{01}(t) \frac{\hat{S}(t-)}{Y_0(t)} dN_{01}(t) = \sum_{i=1}^n e^{-rT_i} y_i \frac{\hat{S}(T_i-)}{Y_0(T_i)} [T_i \leq U_i \wedge \tau].$$

If $\hat{G}$ denotes the Kaplan-Meier estimator of the survival distribution of $U_i$, and using the fact that $\hat{S}(t-)\hat{G}(t-) = n^{-1}Y_0(t)$ if there are no ties between survival and censoring times, then (7) can be rewritten as

$$\text{(8)} \quad \hat{\text{NPV}} = n^{-1} \int_0^\tau e^{-rt} \frac{\hat{c}_{01}(t)}{\hat{G}(t-)} dN_{01}(t) = n^{-1} \sum_{i=1}^n e^{-rT_i} y_i [T_i \leq U_i \wedge \tau]/\hat{G}(T_i-).$$

Using the consistency of the Kaplan-Meier estimator $\hat{G}$ we see that $\hat{\text{NPV}}$ converges to $E(e^{-rT} y[T \leq \tau])$ provided $G(\tau-) > 0$. Therefore in the absence of discounting $\hat{\text{NPV}}$ estimates the average cost restricted to $\tau$. In this case (8) with $r = 0$ is the mean cost estimator described by Bang and Tsiatis [2] and Zhao and Tian [21].

If there are ties in the survival times and $0 < t_1^* < \ldots < t_p^* \leq \tau$ are the distinct observed times, then $\hat{c}_{01}(t_j^*) = \bar{y}_j^*$ is the mean of the observed costs at time $t_j^*$ and the right-hand side of (8) is

$$n^{-1} \sum_{j: t_j^* \leq \tau} e^{-rt_j^*} d_j \bar{y}_j^* / \hat{G}(t_j^*-),$$

where $d_j$ is the multiplicity of $t_j^*$.

### 3.2. Single sojourn without covariates

A single sojourn begins in state 0 and ends with transition to state 1 at time $T$. Sojourn cost is incurred through time $T^* = \min(T, \tau)$. From (6) the NPV of interest is

$$\text{(9)} \quad \text{NPV} = \int_0^\tau e^{-rt} S(t-) b_0(t) dt = \int_0^\tau S(t-) dm(t)$$

where $m(t) = \int_0^t e^{-ru} b_0(u) du$. Allowing for an initial cost at $t = 0$, integration-by-parts yields

$$\text{(10)} \quad \text{NPV} + m(0) = E(m(T^*)) = \int_0^\tau m(t)(-dS(t)) + m(\tau)S(\tau)$$

where $m(0)$ is the expected initial cost. In the absence of discounting ($r = 0$) and ignoring covariates, Strawderman [18] considers the nonparametric estimation of NPV based on observations on (censored) survival times and accumulating costs in $[0, \tau]$. For the $i$-th subject the observed data are $(N_i(t), Y_i(t), V_i(t) : t \leq \tau)$, where $V_i(t)$ is the accumulated costs up to time $t$, $N_i(t) = [T_i \leq t, T_i \leq U_i]$, and $Y_i(t) = [T_i \wedge U_i \geq t]$.

Define $R(t, u) = E(V_i(t) | T_i \geq u)$ for $t \geq u$ and estimate

$$m(t) = \int_0^t E(R(du, u) | T \geq u)$$

by

$$\hat{m}(t) = \sum_{i=1}^{n} \int_0^t \{Y_0(u)\}^{-1} Y_i(u) dV_i(u).$$

This leads to the estimator of NPV,

(11) $$\hat{\text{NPV}} = \sum_{i=1}^{n} \int_0^\tau \hat{S}(t-) Y_i(t) Y_0^{-1}(t) dV_i(t)$$

where, as before $\hat{S}$ is the Kaplan-Meier estimator of $S$ and $Y_0(t) = \sum_{i=1}^{n} Y_i(t)$. To include in (11) a cost at $t = 0$ we could add the term $n^{-1} \sum_{i=1}^{n} \hat{m}_i(0)$ as the estimator of $m(0)$. Because

$$\hat{S}(t) - \hat{S}(t-) = -\hat{S}(t-) \frac{\Delta N_{01}(t)}{Y_0(t)}$$

the right hand side of (10) would be estimated by

$$\sum_{i=1}^{n} \hat{m}(T_i) \hat{S}(T_i-) \{Y_0(T_i)\}^{-1} [T_i \leq U_i \wedge \tau] + \hat{m}(\tau) \hat{S}(\tau).$$

Expression (11) is useful when the accumulating cost history is observed.

### 3.3. Single sojourn without covariates with restricted cost history

Suppose the cost accumulation process $V_i(t)$ is observed at fixed time points $\{a_0, \ldots, a_G\}$ where $0 = a_0 < a_1 < \cdots < a_G = \tau$. Let $V_{ig} = V_i(a_g) - V_i(a_{g-1})$. If observation goes past $a_g$ then $V_{ig}$ is observed. If $T_i \in (a_{g-1}, a_g]$ then $V_i(a_g) = V_i(T_i)$ and if $T_i \leq a_{g-1}$, $V_{ig} = 0$. When censoring occurs in $(a_{g-1}, a_g]$ the true incremental cost in the interval is not known. We only observe $\tilde{V}_{ig} = V_i(U_i) - V_i(a_{g-1})$. In all other cases we define $\tilde{V}_{ig} = V_{ig}$. Regarding $dV_i(t)$ in (11) as a discrete measure with mass $\tilde{V}_{ig}$ at $t = a_{g-1}$ we obtain

(12) $$\hat{\text{NPV}} = \sum_{g=1}^{G} \hat{S}(a_{g-1}-) Y_0^{-1}(a_{g-1}) \sum_{i=1}^{n} Y_i(a_{g-1}) \tilde{V}_{ig}.$$

This estimator was introduced by Lin et al. [15]. By the weak law of large numbers and the independence of $U_i$ with $T_i$ and $V_i(t)$

$$Y_0^{-1}(a_{g-1}) \sum_{i=1}^{n} Y_i(a_{g-1}) V_{ig} \to E(Y_i(a_{g-1}) V_{ig}) / E(Y_i(a_{g-1})) = E(V_{ig} | T_i \geq a_{g-1}).$$

Because $\tilde{V}_{ig}$ differs from $V_{ig}$ when there is censoring, (12) converges to

$$\sum_{g=1}^{G} S_1(a_{g-1}-) E(V_{ig} | T_i \geq a_{g-1}) - E^* = E(V_i(\tau)) - E^*$$

where $E^* = \sum_{g=1}^{G} E\{(V_i(a_g) - V_i(U_i))[U_i \leq T_i \wedge a_g] | U_i \geq a_{g-1}\}$. Hence there is downward bias in estimating the mean cost $E(V_i(\tau))$. If censoring does occur close to the right endpoint of the intervals this bias is likely to be small.

*3.4. Regression model-based estimates of NPV*

For the $i$-th subject let $Y_i = (y_{i1}, \ldots, y_{in_i})'$ denote the costs for the sojourns ending at chronologically ordered times $t_i = (t_{i1}, \ldots, t_{in_i})'$. If the last sojourn has not ended with transition to an absorbing state, we will define $t_{in_i} = \tau$ so that $y_{in_i}$ is the cost associated with the period $[t_{in_i-1}, \tau]$. Censoring would also preclude observation of some sojourn costs. Observation ends in one of three ways: (1) censoring occurs at $U_i$ before $\tau$, (2) an absorbing state is reached before $\tau$, or (3) observation goes past $\tau$. The cost $y_{ig}$ associated with the $g$-th sojourn interval $(t_{ig-1}, t_{ig}]$ is observed if $s_{ig} = 1$ where $s_{ig} = [U_i \geq t_{ig} \wedge \tau]$.

Let $\mathbf{s}_i$ denote the diagonal matrix of the $\{s_{ig}, g = 1, \ldots, n_i\}$ and $\mathbf{X}_i = (\mathbf{x}_{i1}, \ldots, \mathbf{x}_{in_i})'$ be a $n_i \times p$ matrix of covariates associated with $\mathbf{Y}_i$. The components of $x_{ig}$ contain covariates that are fixed over time as well as covariates that vary with time, but only through $(t_{i1}, \ldots, t_{ig})$. In particular $x_{ig}$ will contain functions of $t_{ig-1}, t_{ig}$. The conditional mean vector and covariance matrix are denoted, respectively, by $\boldsymbol{\mu}_i = E(\mathbf{Y}_i|\mathbf{X}_i)$, $\mathbf{V}_i = E[(\mathbf{Y}_i - \boldsymbol{\mu}_i)(\mathbf{Y}_i - \boldsymbol{\mu}_i)'|\mathbf{X}_i]$. We impose *strict exogeneity* on the conditional means $\mu_{ig} = E(y_{ig}|\mathbf{X}_i)$ that requires $\mu_{ig}$ to be a function of $\mathbf{x}_{ig}$ only, that is, $E(y_{ig}|\mathbf{x}_{i1}, \ldots \mathbf{x}_{in_i}) = E(y_{ig}|\mathbf{x}_{ig})$ for all $g = 1, \ldots, n_i$. Independence across subjects is assumed, in fact that $\{(\mathbf{Y}_i, \mathbf{X}_i, \mathbf{s}_i) : 1 \leq i \leq n\}$ is a random sample. The total number of records in the sample is $N = \sum_{i=1}^{n} n_i$.

Let $h$ be a link function such that $h(\mu_{ig}) = \mathbf{x}'_{ig}\beta$ where $\beta$ is a $p \times 1$ vector of unknown parameters. (The $\beta$ here is not the same as the regression parameter in the intensity model (3) of section 2.3.) The $n_i \times p$ matrix $\mathbf{D}_i$ of derivatives $\frac{\partial \boldsymbol{\mu}_i}{\partial \beta'}$ can be expressed as $\mathbf{D}_i = \mathbf{D}_i^0 \mathbf{X}_i$ where $\mathbf{D}_i^0$ is the diagonal matrix with elements $(dh/dx)^{-1}$ evaluated at $x = \mu_{ig}$. Assuming $\mathbf{V}_i$ is positive definite we may write $\mathbf{V}_i = \mathbf{L}_i\mathbf{L}'_i$ where $\mathbf{L}_i$ is the unique lower triangular matrix with positive diagonal elements. Make the transformations $\tilde{\mathbf{Y}}_i = \mathbf{w}_i^{1/2}\mathbf{L}_i^{-1}\mathbf{Y}_i$, $\tilde{\boldsymbol{\mu}}_i = \mathbf{w}_i^{1/2}\mathbf{L}_i^{-1}\boldsymbol{\mu}_i$ where $\mathbf{w}_i$ is the diagonal matrix with elements $w_{ig} = s_{ig}/p(t_{ig} \wedge \tau-, \mathbf{z}_i)$ and $p(t, \mathbf{z}_i) = P[U_i > t|\mathbf{z}_i]$. Here $\mathbf{z}_i$ are fixed covariates that model the censoring distribution. They may differ from the components of $\mathbf{X}_i$. Given $\mathbf{z}_i$, assume $U_i$ is independent of $(\mathbf{Y}_i, \mathbf{X}_i, \mathbf{t}_i)$. Then $E(s_{ig}|\mathbf{Y}_i, \mathbf{X}_i, \mathbf{t}_i, \mathbf{z}_i) = P[U_i \geq t_{ig} \wedge \tau|\mathbf{z}_i]$ and $E(w_{ig}|\mathbf{Y}_i, \mathbf{X}_i, \mathbf{t}_i, \mathbf{z}_i) = 1$ under the assumption $p(\tau-, \mathbf{z}_i) > 0$.

An estimator of $\beta$ is obtained by minimizing the sum of squares $\tilde{q}(\mathbf{Y}_i, \mathbf{w}_i, \mathbf{X}_i) = \sum_{i=1}^{n}(\tilde{\mathbf{Y}}_i - \tilde{\boldsymbol{\mu}}_i)'(\tilde{\mathbf{Y}}_i - \tilde{\boldsymbol{\mu}}_i)$ with respect to $\beta$ which leads to the estimating equation

$$(13) \qquad \sum_{i=1}^{n} \mathbf{D}'_i(\mathbf{L}_i^{-1})'\mathbf{w}_i(\mathbf{L}_i^{-1})(\mathbf{Y}_i - \boldsymbol{\mu}_i) = 0.$$

Because

$$E[\mathbf{D}'_i(\mathbf{L}_i^{-1})'\mathbf{w}_i(\mathbf{L}_i^{-1})(\mathbf{Y}_i - \boldsymbol{\mu}_i)]$$
$$= E[\mathbf{D}'_i(\mathbf{L}_i^{-1})'E(\mathbf{w}_i|\mathbf{Y}_i, \mathbf{X}_i, \mathbf{t}_i, \mathbf{z}_i)(\mathbf{L}_i^{-1})(\mathbf{Y}_i - \boldsymbol{\mu}_i)]$$
$$= E[\mathbf{D}'_i\mathbf{V}_i^{-1}(\mathbf{Y}_i - \boldsymbol{\mu}_i)] = 0,$$

(13) provides a consistent estimator $\hat{\beta}$ of $\beta$. The transformation of $\mathbf{Y}_i - \boldsymbol{\mu}_i$ and $\mathbf{D}_i$ by $\mathbf{w}_i^{1/2}\mathbf{L}_i^{-1}$ preserves time order and effectively uses only uncensored data in (13). In the absence of censoring we would use the estimating equation

$$\sum_{i=1}^{n} \mathbf{D}'_i\mathbf{V}_i^{-1}(\mathbf{Y}_i - \boldsymbol{\mu}_i) = 0.$$

Hence (13) is the generalized estimating equations (GEE) analog for the selected sample $\{(\mathbf{Y}_i, \mathbf{X}_i, \mathbf{s}_i) : 1 \leq i \leq n\}$.

Following the standard GEE methodology, $n^{1/2}(\hat{\beta} - \beta)$ is asymptotically normal with zero mean and covariance matrix $\mathbf{A}^{-1}\mathbf{B}\mathbf{A}^{-1}$ where

$$
(14) \quad \begin{aligned}
\mathbf{A} &= E\left(\frac{\partial \mathbf{S}_i(\mathbf{w}_i, \mathbf{Y}_i, \mathbf{X}_i, \beta)}{\partial \beta'}\right), \\
\mathbf{B} &= E[\mathbf{S}_i(\mathbf{w}_i, \mathbf{Y}_i, \mathbf{X}_i, \beta)\mathbf{S}'_i(\mathbf{w}_i, \mathbf{Y}_i, \mathbf{X}_i, \beta)]
\end{aligned}
$$

and $\mathbf{S}_i(\mathbf{w}_i, \mathbf{Y}_i, \mathbf{X}_i, \beta) = \mathbf{D}'_i(\mathbf{L}_i^{-1})'\mathbf{w}_i(\mathbf{L}_i^{-1})(\mathbf{Y}_i - \boldsymbol{\mu}_i)$. Consistent estimators of $\mathbf{A}$ and $\mathbf{B}$ are obtained by replacing the expectations in (14) by their sample averages and $\beta$ by $\hat{\beta}$. In addition, we also need a consistent estimator of $\mathbf{V}_i = \mathbf{L}_i\mathbf{L}'_i$ and the censoring distribution $p(t, \mathbf{z}_i)$. Methods for their estimation are suggested in specific contexts in Lin [13, 14], Baser et al. [5] and Gardiner et al. [8].

Another approach is to estimate a random-effects (RE) model for $\mathbf{Y}_i$ (or a transformation of $\mathbf{Y}_i$) given by

$$(15) \quad \mathbf{Y}_i = \mathbf{X}_i\beta + a_i\mathbf{1}_i + \mathbf{u}_i$$

where $\beta$ is an unknown $p \times 1$ parameter, $\mathbf{1}_i$ the $n_i \times 1$ vector with all elements equal to 1, $a_i$ an unobserved patient-specific heterogeneity and $\mathbf{u}_i$ is the $n_i \times 1$ vector of idiosyncratic errors. The composite error is $\mathbf{v}_i = a_i\mathbf{1}_i + \mathbf{u}_i$. Assume $\boldsymbol{\Omega}_i = E(\mathbf{v}_i\mathbf{v}'_i)$ is positive definite and that the standard RE assumptions (Wooldridge [20]) hold:

(a) $E(\mathbf{u}_i|\mathbf{X}_i, a_i) = \mathbf{0}$, $E(a_i|\mathbf{X}_i) = 0$,
(b) rank $E(\mathbf{X}'_i\boldsymbol{\Omega}_i^{-1}\mathbf{X}_i) = p$,
(c) $E(\mathbf{u}_i\mathbf{u}'_i|\mathbf{X}_i, a_i) = \sigma_u^2 \mathbf{I}_i$, $E(a_i^2|\mathbf{X}_i) = \sigma_a^2$ where $\sigma_u^2$ and $\sigma_a^2$ are constants and $\mathbf{I}_i$ is the $n_i \times n_i$ identity matrix. Therefore $E(\mathbf{v}_i) = \mathbf{0}$ and $\boldsymbol{\Omega}_i = \sigma_u^2 \mathbf{I}_i + \sigma_a^2 \mathbf{J}_i$ where $\mathbf{J}_i$ is the $n_i \times n_i$ matrix with all elements equal to 1.

To estimate $\beta$ in (15) from censored observations on costs we first transform $(\mathbf{Y}_i, \mathbf{X}_i, \mathbf{v}_i)$ to $(\tilde{\mathbf{Y}}_i, \tilde{\mathbf{X}}_i, \tilde{\mathbf{v}}_i)$ where $\tilde{\mathbf{v}}_i = \mathbf{w}_i^{1/2}\mathbf{L}_i^{-1}\mathbf{v}_i$ and $\tilde{\mathbf{Y}}_i, \tilde{\mathbf{X}}_i$ are similarly defined. Here $\mathbf{L}_i$ is the unique lower triangular matrix with positive diagonal elements such that $\boldsymbol{\Omega}_i = \mathbf{L}_i\mathbf{L}'_i$. The objective function for estimating $\beta$ is

$$\tilde{q}(\mathbf{Y}_i, \mathbf{w}_i, \mathbf{X}_i) = \{\mathbf{w}_i^{1/2}(\mathbf{L}'_i)^{-1}(\mathbf{Y}_i - \mathbf{X}_i\beta)\}'\{\mathbf{w}_i^{1/2}\mathbf{L}_i^{-1}(\mathbf{Y}_i - \mathbf{X}_i\beta)\}.$$

Specializing (13) leads to the generalized least-squares (GLS) weighted estimator $\hat{\beta}_w$ given by

$$(16) \quad \hat{\beta}_w = \left(\sum_{i=1}^n \tilde{\mathbf{X}}'_i\tilde{\mathbf{X}}_i\right)^{-1}\left(\sum_{i=1}^n \tilde{\mathbf{X}}'_i\tilde{\mathbf{Y}}_i\right).$$

From (16) we get the consistency of $\hat{\beta}_w$ and

$$(17) \quad n^{1/2}(\hat{\beta}_w - \beta) \to N(0, \mathbf{A}^{-1}\mathbf{B}\mathbf{A}^{-1})$$

where $\mathbf{A} = E(\mathbf{X}'_i\boldsymbol{\Omega}_i^{-1}\mathbf{X}_i)$ and $\mathbf{B} = E(\tilde{\mathbf{X}}'_i\tilde{\mathbf{v}}_i\tilde{\mathbf{v}}'_i\tilde{\mathbf{X}}_i)$.

### 3.5. Estimation of NPV

From our model (15) for all transition costs we obtain estimates of $c_{hj}(t|\mathbf{z})$ for a covariate profile $\mathbf{z}$ by specifying the covariates $\mathbf{x}_0$ corresponding to column positions in **X**. The row vector $\mathbf{x}'_{ij}$ of $\mathbf{X}_i$ in our model for $y_{ij}$ will contain the fixed covariates $\mathbf{x}_i$, dummies for transitions types, terms of modeling the transition times such as $t_{ij}$, $t_{ij}^2$ and perhaps interactions between these times and $\mathbf{x}_i$. Our special $\mathbf{x}_0$ will contain the desired **z**, interactions between **z**, $t$ and $t^2$, indicator variables with value 1 for transition type $h \to j$, and value 0 for all other transition types. Denoting this covariate profile by $\mathbf{x}_{hj0}(t)$ then $c_{hj}(t|\mathbf{z}) = x'_{hj0}(t)\beta$ and from (16) we obtain the estimator

$$\hat{c}_{hj}(t|\mathbf{z}) = \mathbf{x}'_{hj0}(t)\hat{\beta}_w. \tag{18}$$

Although the consistency of $\hat{c}_{hj}(t|\mathbf{z})$ might seem immediate from (18) the final form of the computable $\hat{\beta}_w$ involves the estimated $\mathbf{\Omega}_i$ and weights $\mathbf{w}_i$, the latter through the censoring distribution $G$. A formal verification is not attempted here, but see Baser et al. [5] for a similar context.

Now recall the expected net present value $E(C_{hj}^{(1)}|X(0) = i, \mathbf{z})$. Plugging in estimators for the entities on the right hand side in (4) leads to

$$\hat{E}(C_{hj}^{(1)}(\tau)|X_0 = i, \mathbf{z}) = \int_0^\tau e^{-rt}\hat{c}_{hj}(t|\mathbf{z})\hat{P}_{ih}(0, t-|\mathbf{z})d\hat{A}_{hj}(t|\mathbf{z}). \tag{19}$$

The estimation of $E(C_h^{(2)}|X(0) = i, \mathbf{z})$ is entirely analogous except that one must deal with the quantity $b_h(t|\mathbf{z})$ which is the expected mean rate of expenditures at time $t$ while sojourning in state $h$. In practice it will not be observable unless discrete information is available. Instead, we will know only the total cost of the sojourn. For example, consider hospital costs for patients undergoing coronary artery bypass surgery. Expenditures are incurred in various care units such as the intensive care unit, cardiac care unit and in recovery. We would know the entry and exit dates for each unit and the associated cost of the length of stay in each unit, but not necessarily the cost per day. An application modeling treatment cost rates in cancer patients is discussed in Gardiner et al. [8] using a model for the log-transformed rate of cost accumulation $y_{ij} = y_{ij}^*/(t_{ij} - t_{ij-1})$ between consecutive transition times $t_{i1}, t_{i2}, \ldots$ where $y_{ij}^*$ the sojourn cost in $[t_{ij-1}, t_{ij})$.

### 3.6. Single transition with covariates

Consider the same scenario discussed previously in 3.1 with all patients starting in state "0" and followed until they reach the terminal state "1" (dead). For the $i$-th patient $T_i$ is the survival time and $U_i$ the censoring time. Observation ceases at $\min(T_i, U_i, \tau)$, that is, either at the failure time, or censoring time or the limit of observation. The only cost incurred is $y_i = y_i(T_i)$ at time $T_i$ which is observed if $s_i = 1$ where $s_i = [U_i \wedge \tau \geq T_i]$. Let $\mathbf{x}_i$ denote a $p$-vector of fixed covariates of interest and $\mathbf{z}_i$ denote fixed covariates used for modeling the censoring distribution. Assuming independent censoring, that is, given $\mathbf{z}_i$, $U_i$ is independent of $(y_i, \mathbf{x}_i, T_i)$, we get

$$P[s_i = 1|\mathbf{z}_i, y_i, \mathbf{x}_i, T_i] = P[U_i \geq T_i, T_i \leq \tau|\mathbf{z}_i, T_i] = G(T_i-|\mathbf{z}_i)[T_i \leq \tau].$$

Defining $w_i = s_i/G(T_i-|\mathbf{z}_i)$ we see that (13) reduces to minimizing with respect to $\beta$ the objective function $n^{-1}\sum_{i=1}^n q(y_i, w_i, \mathbf{x}_i)$ where $q(y_i, w_i, \mathbf{x}_i) = \sigma_u^{-2}w_i(y_i - \mathbf{x}'_i\beta)^2$. This yields the estimator $\hat{\beta}_w$ in (16) which in this case is

$$\hat{\beta}_w = \left(\sum_{i=1}^n w_i\mathbf{x}_i\mathbf{x}'_i\right)^{-1}\sum_{i=1}^n w_i\mathbf{x}_iy_i.$$

This is the same estimator described by Lin [13] except for a slight difference in the weights. Because Lin [13] uses a model in which costs are incurred through time $T_i \wedge \tau$ his censoring indicator is $s_i^* = [U_i \geq T_i \wedge \tau]$ and weight $w_i^* = s_i^*/G(T_i \wedge \tau - |\mathbf{z}_i)$. In our case the cost is realized at time $T_i$ if and only if $s_i = 1$.

Let $\mathbf{z}_0$ denote a fixed covariate at which NPV($\mathbf{z}_0$) is to be estimated. Since "0" is the initial state and the only transition is 0→1, $\hat{P}_{00}(0, t - |\mathbf{z}_0) = \hat{S}(t - |\mathbf{z}_0)$ and $\hat{S}(t|\mathbf{z}_0) = \exp(-\hat{A}_{01}(t|\mathbf{z}_0))$. Here $\hat{S}$ is the estimator of the survival distribution $S$ of $T$, the time of transition. From (18) and (19) our estimator of NPV($\mathbf{z}_0$) is

$$\widehat{\text{NPV}}(\mathbf{z}_0) = \hat{\beta}'_w \int_0^\tau e^{-rt} \mathbf{x}_0(t)\{-d\hat{S}(t|\mathbf{z}_0)\},$$

where $\mathbf{x}_0(t)$ is the covariate vector derived from $\mathbf{z}_0$ and terms used to model time (such as $t, t^2$) in the cost equation $y_i = \mathbf{x}'_i \beta + u_i$. Here a single cost $y_i = y_i(T_i)$ is incurred at $T_i$, if observed by time $\tau$. Then

$$E[y_i(t)|\mathbf{z}_0, T_i = t] = \mathbf{x}'_0(t)\beta$$

and

$$\text{NPV}(\mathbf{z}_0) = \int_0^\tau e^{-rt} E[y_i(t)|\mathbf{z}_0, T_i = t]\{-dS(t|\mathbf{z}_0)\}$$

simplifies to $E(e^{-rT_i} y_i(T_i)[T_i \leq \tau]|\mathbf{z}_0)$.

Since $\hat{\beta}_w \to \beta_w$ in probability, and uniformly on $[0, \tau]$, $\hat{S}(\cdot|z_0) \to S(\cdot|\mathbf{z}_0)$ in probability, if $S(\tau|\mathbf{z}_0) > 0$, we obtain the consistency of $\widehat{\text{NPV}}(\mathbf{z}_0)$ provided $\int_0^\tau e^{-rt} \mathbf{x}_0(t) dS(t|\mathbf{z}_0)$ is finite. Also in estimating $\beta_w$ we require

$$P[T_i \leq \tau|\mathbf{z}_0] = 1 - S(t|\mathbf{z}_0) > 0,$$

because otherwise the cost equation will be vacuous since with probability 1 no transition takes place in $[0, \tau]$.

### 3.7. Single sojourn with covariates

Suppose the interval $[0,\tau]$ is partitioned by the fixed points $a_j, j = 0, \ldots, K$ with $0 = a_0 < a_1 < \ldots < a_K = \tau$. If the expected rate of cost accumulation is constant in the intervals $(a_{j-1}, a_j)$ with values $b_j$ we have

$$\text{NPV}(\mathbf{z}_0) = \sum_{j=1}^K b_j \int_{a_{j-1}}^{a_j} e^{-rt} S(t|\mathbf{z}_0) dt.$$

The integral is the increment over $(a_{j-1}, a_j)$ in discounted life expectancy,

$$LE(\mathbf{z}_0, t) = \int_0^t e^{-ru} S(t|\mathbf{z}_0) du.$$

Following Baser et al. [5] we could use a RE model for cost $\mathbf{y}_i = (y_{i1}, \ldots, y_{iK})'$ incurred by the $i$-th patient. Here $y_{ij}$ is the cost incurred in interval $(a_{j-1}, a_j)$ which is observed provided $s_{ij} = 1$ where $s_{ij} = [T_i \wedge U_i \geq a_j] + [a_{j-1} < T_i < U_i \wedge a_j]$. This reflects the two cases: (1) the patient neither died nor was censored before $a_j$, or (2) death was observed in $(a_{j-1}, a_j)$. Under the assumed independence of censoring $P[s_{ij} = 1|\mathbf{z}_i, T_i] = G(T_{ij}^* - |\mathbf{z}_i)[T_i \geq a_{j-1}]$ where $T_{ij}^* = \min(T_i, a_j)$. The regression model for $y_{ij}$ will include interval-specific elapsed time $T_{ij}^* - a_{j-1}$ which will yield an estimator of $b_j$ that may depend on $\mathbf{z}_0$.

## 4. Discussion and summary

The estimation of medical costs has received considerable attention because of its importance in assessing cost-effectiveness of medical interventions and treatments. Facing constrained healthcare budgets government planners and policy makers are forced to consider the costs of competing interventions in addition to claims of their clinical efficacy. The difference in the expected cost of two competing interventions is the numerator of the cost-effectiveness ratio, the denominator being the incremental health benefit as measured by life expectancy or by quality-adjusted life years (Chen and Sen [6, 7] and Gardiner et al. [9]). The cost-effectiveness ratio can be used to compare competing interventions with respect to both their health benefits as well as their cost.

Obtaining reliable and valid estimates of costs is imperative. In this article we adopted a longitudinal framework in which patient costs are manifested dynamically over time. An underlying finite-state stochastic process describes the evolving patient history with costs incurred at transition times between states and during sojourn in states. In this framework we showed how net present values are defined, following the basic notions of actuarial values used extensively in the insurance and finance literature (Norberg [16]). For example, in the classic disability model there are "able" periods and "disabled" periods. The individual holding a disability insurance policy would receive a fixed payment stream over the period of his or her disability. In able periods the individual would pay the fixed premium in accordance with the policy. There are three policy states –"able", "disabled", and "dead". A fundamental difference in our context is that costs are not fixed but random. One may regard the total cost over a specified period as the sum of all transition costs and sojourn costs.

Several methods have been proposed to estimate medical cost from follow up data. The primary focus has been on a single cost measure that might be incompletely ascertained due to time censoring (Bang and Tsiatis [2], Baser et al. [4], Lin et al. [15], Strawderman [18] and O'Hagan and Stevens [17]). Regression analyses allow for assessing the influence of explanatory variables on some measure of the cost distribution, such as the mean or median (Bang and Tsiatis [3], Baser et al. [5], Lin [13, 14] and Gardiner et al. [8]). Apart from addressing the incompleteness of cost data, the ability to observe costs over finer time periods can serve to strengthen ensuing analyses. For instance, consider the cost of a treatment which is assumed to last at most for one year and costs are monitored monthly. If either the endpoint is reached before the end of the year, or observation lasts one year, the total cost is observed. If there is censoring of the endpoint before the end of the year, we could use the monthly costs, except for the last month of observation, to improve our estimate of the average cost of treatment.

The methods discussed here for analyses of medical costs may be adapted to estimate other summary measures used in cost-effectiveness analyses (Gardiner et al. [9]). For example, quality-adjusted survival is defined by using a quality weight $q(h,t)$ which represents the utility, relative to the state of perfect health, of each unit of time spent in state $h = X(t)$ at time $t$. Perfect health has a quality weight 1, while death or states judged equivalent to death get a quality weight of 0. *The total quality adjusted time* in $[0, \tau]$ is $\sum_{h \in E} \int_0^\tau e^{-rt} q(h,t) Y_h(t) dt$. Hence conditional on $X(0) = i$ we define the *expected quality adjusted life years*, $QALY_i(\mathbf{z}) = \sum_{h \in E} \int_0^\tau e^{-rt} q(h,t) P_{ih}(0, t-|\mathbf{z}) dt$. The unconditional version is given by $QALY(\mathbf{z}) = \sum_{i \in E} \pi_i(0|\mathbf{z}) QALY_i(\mathbf{z})$. This expression is similar to the second term of the NPV in (6).

The transition model adopted in this article extends the simpler two-state survival model with a single transition and sojourn. The underlying analysis of survival times is now replaced by the analysis of multiple event times which is facilitated by using a non-homogeneous Markov model to govern the movement between states, and a multiplicative intensity model to incorporate covariate effects. For the analysis of longitudinal cost data, techniques such as inverse probability weighting to account for censoring can be applied (Willan et al. [19]) but a more careful consideration is required to combine the two parts of the model, the transition model for the event times and a regression model for costs. Methods for joint modeling of longitudinal observations and event times could be adapted for this purpose (Henderson et al. [10] and Hogan and Laird [11, 12]).

**Acknowledgments.** We thank the referees for their careful reading of the manuscript. Their comments have improved the presentation of the paper.

## References

[1] ANDERSEN, P. K., BORGAN, O., GILL, R. D. AND KEIDING, N. (1993). *Statistical Models Based on Counting Processes*. Springer, New York. MR1198884
[2] BANG, H. AND TSIATIS, A. A. (2000). Estimating medical costs with censored data. *Biometrika* **87** 329–343. MR1782482
[3] BANG, H. AND TSIATIS, A. A. (2002). Median regression with censored cost data. *Biometrics* **58** 643–649. MR1926117
[4] BASER, O., GARDINER, J. C., BRADLEY, C. J. AND GIVEN, C. W. (2004). Estimation from censored medical cost data. *Biometrical Journal* **46** 351–363. MR2062000
[5] BASER, O., GARDINER, J. C., BRADLEY, C. J., YUCE, H. AND GIVEN, C. (2006). Longitudinal analysis of censored medical cost data. *Health Economics* **15** 513–525.
[6] CHEN, P. L. AND SEN, P. K. (2001). Quality-adjusted survival estimation with periodic observations. *Biometrics* **57** 868–874. MR1863451
[7] CHEN, P. L. AND SEN, P. K. (2004). Quality-adjusted survival estimation with periodic observations: A multistate survival analysis approach. *Comm. Statist. Theory Methods* **33** 1327–1339. MR2069571
[8] GARDINER, J. C., LUO, Z., BRADLEY, C. J., SIRBU, C. M. AND GIVEN, C. W. (2006a). A dynamic model for estimating changes in health status and costs. *Statistics in Medicine* **25** 3648–3667. MR2252417
[9] GARDINER, J. C., LUO, Z., LIU, L. AND BRADLEY, C. J. (2006b). A stochastic framework for estimation of summary measures in cost-effectiveness analyses. *Expert Review of Pharmacoeconomics and Outcomes Research* **6** 347–358.
[10] HENDERSON, R., DIGGLE, P. AND DOBSON, A. (2000). Joint modelling of longitudinal measurements and event time data. *Biostatistics* **1** 465–480.
[11] HOGAN, J. W. AND LAIRD, N. M. (1997a). Mixture models for the joint distribution of repeated measures and event times. *Statistics in Medicine* **16** 239–257.
[12] HOGAN, J. W. AND LAIRD, N. M. (1997b). Model-based approaches to analysing incomplete longitudinal and failure time data. *Statistics in Medicine* **16** 259–272.
[13] LIN, D. Y. (2000). Linear regression of censored medical costs. *Biostatistics* **1** 35–47.

[14] Lin, D. Y. (2003). Regression analysis of incomplete medical cost data. *Statistics in Medicine* **22** 1181–1200.
[15] Lin, D. Y., Feuer, E. J., Etzioni, R. and Wax, Y. (1997). Estimating medical costs from incomplete follow-up data. *Biometrics* **53** 419–434.
[16] Norberg, R. (1995). Differential-equations for moments of present values in life-insurance. *Insurance Mathematics and Economics* **17** 171–180. MR1363647
[17] O'Hagan, A. and Stevens, J. W. (2004). On estimators of medical costs with censored data. *Journal of Health Economics* **23** 615–625.
[18] Strawderman, R. L. (2000). Estimating the mean of an increasing stochastic process at a censored stopping time. *Journal of the American Statistical Association* **95** 1192–1208. MR1804243
[19] Willan, A. R., Lin, D. Y., Cook, R. J. and Chen, E. B. (2002). Using inverse-weighting in cost-effectiveness analysis with censored data. *Statistical Methods in Medical Research* **11** 539–551.
[20] Wooldridge, J. M. (2002). *Econometric Analysis of Cross Section and Panel Data*. MIT Press, Cambridge, MA.
[21] Zhao, H. W. and Tian, L. L. (2001). On estimating medical cost and incremental cost-effectiveness ratios with censored data. *Biometrics* **57** 1002–1008. MR1950417